\documentclass[twocolumn,shortnote]{jpsj3}
\usepackage{graphicx}
\usepackage{amsmath,amssymb,braket,amsfonts}
\usepackage{dcolumn}
\usepackage{bm}
\usepackage{color}

\title{Spin dynamics simulation of the $Z_2$-vortex fluctuations}

\author{
Yo P. Mizuta$^{1}$, Kazuki Aoyama$^{2}$, Keisuke Tomiyasu$^{3}$, Masato Matsuura$^{4}$, and Hikaru Kawamura$^{5}$\thanks{E-mail:h.kawamura.handai@gmail.com}
}

\inst{
$^1$Graduate School of Engineering science, Osaka University, Toyonaka, Osaka 560-8531, Japan\\
$^2$Department of Earth and Space Science, Graduate School of Science, Osaka University, Toyonaka, Osaka 560-0043, Japan\\
$^3$Nissan ARC Ltd., Natsushima-cho 1, Yokosuka 237-0061, Japan\\
$^4$CROSS Neutron Science and Technology Center, IQBRC Bldg, Tokai, Ibaraki 319-1106, Japan\\
$^5$Molecular Photoscience Research Center, Kobe University, Kobe, Hyogo 657-8501, Japan
}

\begin{document}
\maketitle

Motivated by the recent quasi-elastic neutron scattering experiment, we extend the spin-dynamics simulation on the triangular-lattice Heisenberg antiferromagnet, to observe a sharp central peak of its energy width $\sim 0.001J$ ($J$ the exchange coupling) of the $Z_2$-vortex origin, consistently with the experiment.
\bigskip\bigskip

Some time ago, Kawamura and Miyashita pointed out that the frustrated isotropic Heisenberg magnets in two dimensions could possess a novel vortex characterized by the parity-like two-valued topological number corresponding only to its presence/absence, a $Z_2$ vortex, which drove a topological transition at a finite temperature $T=T_V$ associated with its binding-unbinding \cite{KM}. Interestingly, the spin correlation length stays finite even at and below $T_V$, and the low-$T$ phase is the spin paramagnetic state with topologically broken ergodicity, called the spin-gel state \cite{KawamuraYamamoto,Kawamura-review}.

 While the direct experimental observation of the $Z_2$ vortex has remained elusive for years, a very recent quasi-elastic neutron scattering (QENS) experiment  performed on the powder sample of the quasi-two-dimensional (2D) triangular-lattice Heisenberg antiferromagnet NaCrO$_2$ has succeeded in directly probing the signature of the free Z$_2$ vortex via the observation of a sharp quasi-elastic (QE) scattering  of its energy width as narrow as $\sim 10 \mu$eV corresponding to $\sim 0.001J$ ($J$ the exchange coupling) in the finite-$T$ range of 30-40 K \cite{Tomiyasu}. As this material has been considered to be a promising candidate of the $Z_2$-vortex-bearing system \cite{Olariu}, it seems fully consistent to identify the origin of the observed sharp QE scattering as the dynamics of the free $Z_2$ vortex. 

 The theoretical proposal that the signature of the free $Z_2$ vortex might be detectable via the appearance of a sharp central peak (QE scattering) in the dynamical spin structure factor at $T\gtrsim T_V$ was made earlier \cite{OkuboKawamura}. Indeed,  by the spin-dynamics simulation on the triangular-lattice Heisenberg model with the nearest-neighbor (nn) antiferromagnetic (AF) coupling $J$, the appearance of such $Z_2$-vortex-induced central peak of the width $\sim 0.01J$ at $T>T_V\simeq 0.285J$ was numerically obtained \cite{OkuboKawamura}. The central-peak width in the recent experiment on NaCrO$_2$, however, was even narrower than that in the model simulation \cite{OkuboKawamura} by an order of magnitude. One might wonder if it might be possible to numerically reproduce an order-of-magnitide sharper central peak as observed in the recent experiment. \cite{Tomiyasu}

 In order to examine this issue, we extend the earlier simulation of Ref.[\citen{OkuboKawamura}] to longer simulation time and to larger lattices. Longer simulation time is implemented to improve the $\omega$-resolution required to detect the sharp structure as a function of $\omega$, and we implement here twenty times longer simulation time than that of Ref.[\citen{OkuboKawamura}]. Larger lattice sizes enables us to deal with more isolated, long-lived free $Z_2$ vortex, and we treat here the lattices of its linear size twice as large as that treated in Ref.[\citen{OkuboKawamura}].

 The model considered is the same as that in Ref.[\citen{OkuboKawamura}], i.e., the AF classical Heisenberg model on the 2D triangular lattice, whose Hamiltoanin is given by ${\cal H} = J \sum_{i,j}\bm{S}_i\cdot \bm{S}_j$ ($J > 0$), where the sum is taken over all nn pairs on the $L\times L$ triangular lattice under periodic boundary conditions. We set here $L=768$ and 1536.

 Following Ref.[\citen{OkuboKawamura}], the spin dynamics simulation is performed according to the classical analogue of the Bloch equation. The temperature effect is taken into account via the initial spin configurations generated by the equilibrium Monte Carlo (MC) simulation at temperature $T$ based on the combined heat-bath and over-relaxation methods. The time evolution of the spins is solved by the fourth-order Runge-Kutta method, where the time mesh $\Delta t$ is taken to be 0.01 (in the $\hbar=1$ unit), commonly with Ref.[\citen{OkuboKawamura}]. The maximum simulation time $t_{max}$ is taken to be 16,000, twenty times longer than $t_{max}$ of Ref.[\citen{OkuboKawamura}]. Thermal average is taken by averaging over 200-1000 independent runs with different spin initial conditions.

 We compute the dynamical spin structure factor,
\begin{eqnarray}
S(\bm{q}, \omega) &=& \langle |\bm{S}_{\bm{q}}(\omega)|^2 \rangle, \\ 
\bm{S}_{\bm{q}}(\omega) &=& \int {\rm d}t \sum_i \bm{S}_i(t) \exp [-i(\bm{q}\cdot \bm{r}_i+\omega t)], 
\end{eqnarray}
where $\bm{q}$ is the wavevector, $\omega$ the angular frequency, and $\langle \cdots \rangle$ denotes the thermal average.

 Typical $\omega$-dependence of the computed $S(\bm{q}, \omega)$ close to the $K$ point is shown in Fig. 1(a), where the temperature $T/J$=0.295 is slightly above $T_V/J$=0.285 and the wave vector $\bm{q}$ is slightly away from the $K$ point in the direction of the $M$ point with $|\bm{q}-\bm{q}_K|=\frac{2\pi}{192}$ (in the unit of $\frac{1}{a}$, $a$ being the lattice constant). As can be seen from Fig. 1(a), a central peak is observed in addition to the side peak originating from the damped spin waves. The data are collected for various temperatures and wavevectors lying on the line connecting the $K$ and $M$ points, and are fitted by the form,

%
\begin{equation}
C_L \frac{1}{\omega^2+\Gamma_L^2} + C_{{\rm DHO}}\frac{\omega_0^2\gamma}{(\omega^2-\omega_0^2)^2+\Gamma^2\omega^2} +C_0 ,
\end{equation}
where the first Lorentzian term represents the QE central peak of the amplitude $C_L$ and the width $\Gamma_L$, the second term the contribution of the damped harmonic oscillator (DHO) of the frequency $\omega_0$, the width $\Gamma$ and the amplitude $C_{{\rm DHO}}$, the last term being the constant background.  We show in Fig. 1(a) the fitting results of the raw $S(\bm{q}, \omega)$ data where the best fitted curve is shown together with the contribution of the three terms in Eq. (3). Reasonably good fit is obtained.

\begin{figure}[t]
\begin{center}
	\includegraphics[width=0.9\linewidth]{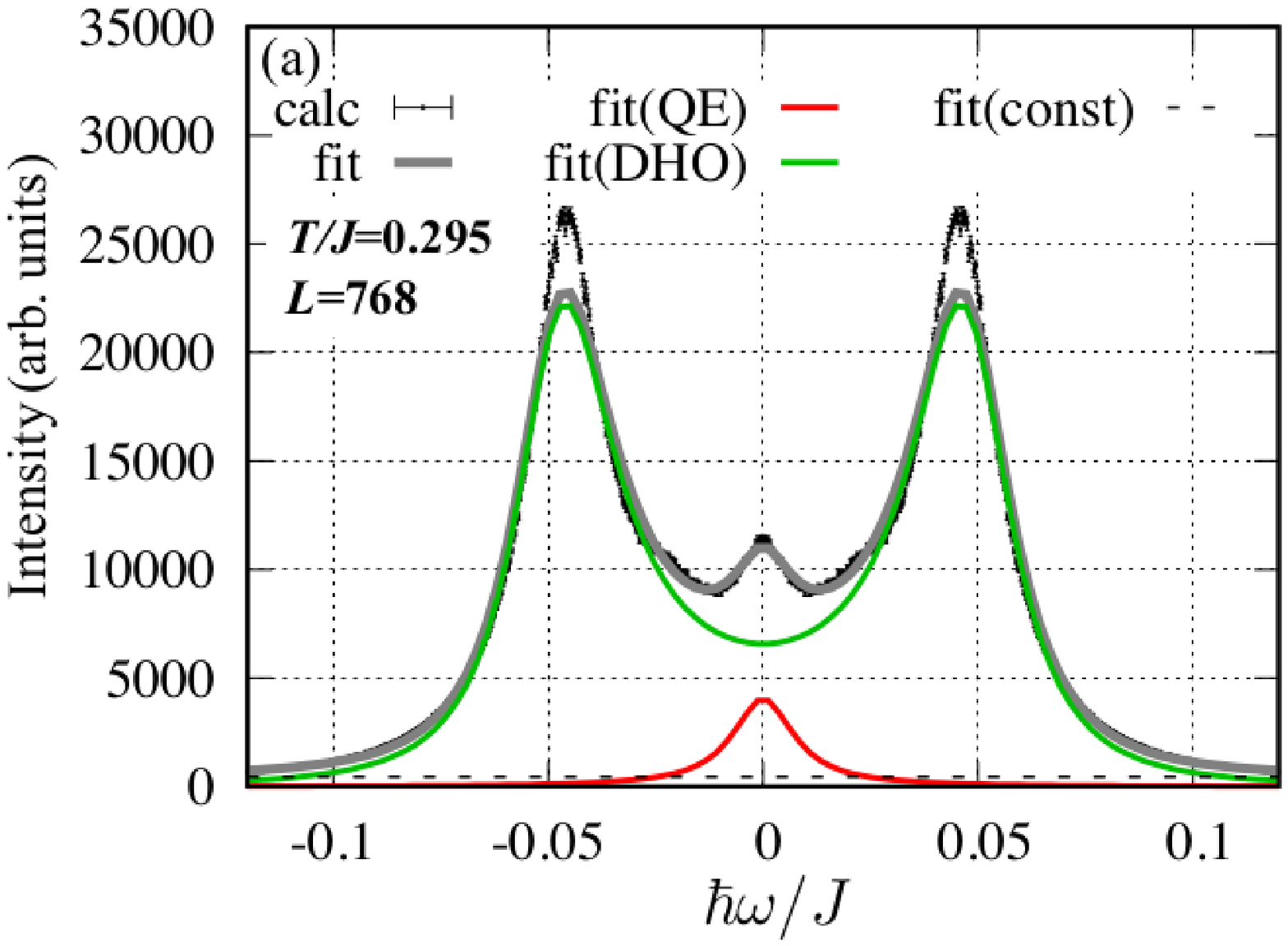}
	\includegraphics[width=0.9\linewidth]{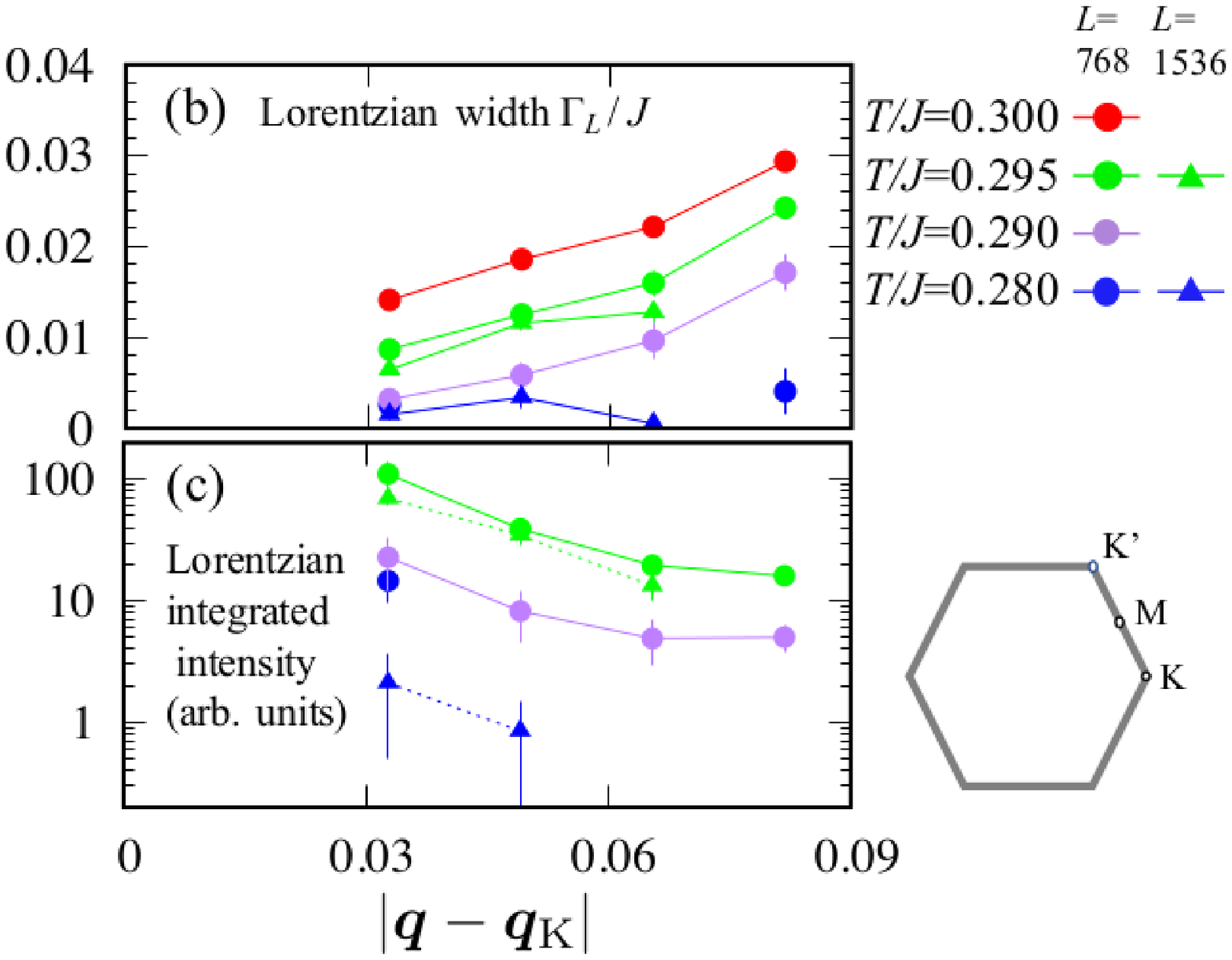}
\end{center}
	\caption{(Color online) (a) The $\omega$ dependence of $S(\bm{q},\omega)$ computed at a temperature $T/J=0.295$ just above $T_V/J=0.285$, and at the wavevector $\bm{q}=\bm{q}_{\rm K}+\frac{1}{64}(\bm{q}_{\rm K}-\bm{q}_{\rm M})$ close to the $K$-point for the size $L=768$. The best fitted curve based on Eq. (3) is also shown, together with the contribution of each term, i.e., the QE central peak, the DHO term, and the constant (const) term. (b) The energy width and (c) the integrated intensity of the fitted QE central peak plotted versus the distance from the $K$ point along the $KM$ line, $|\bm{q}-\bm{q}_{\rm K}|$ in the unit of $\frac{1}{a}$, illustrated in the hexagon representing the 1st Brillouin zone shown at the right bottom.
	} 
\label{fig}
\end{figure}

 Figures 1(b) and 1(c) exhibit the width $\Gamma_L$ (b) and the integrated intensity (c) of the QE central peak plotted versus the wavevector $|\bm{q}-\bm{q}_K|$ for several temeratures around $T_V$. As can be seen from Fig. 1(b), the width $\Gamma_L$ gets narrower as $T$ approaches $T_V$ from above, and as $\bm{q}$ approaches $\bm{q}_K$. In particular, $\Gamma_L$ gets as small as $\sim 0.001J$ just above $T_V$ in the close vicinity of the $K$ point. The earlier estimate $\sim 0.01J$ was obtained a bit away from the $K$ point with the lower $\omega$-resolution \cite{OkuboKawamura}, but is numerically consistent with our present result. One can also see from Fig. 1(c) that the QE central-peak intensity tends to decrease as $T$ approaches $T_V$, eventually vanishing at $T<T_V$, while it gets larger as $\bm{q}$ approaches $\bm{q}_K$. 
 It was already pointed out in Ref.[\citen{OkuboKawamura}] that the width of the QE central peak is quite narrow in the vicinity of the $K$ point and is much boarder far away from the $K$ point, where the former was  associated with the free $Z_2$ vortices, while the latter associated with the bound $Z_2$-vortex pairs. Our present observation, which reveals the systematic variation of the width against $\bm{q}$, is fully consistent with such a previously proposed picture. 

 On the basis of our present observation, we wish to discuss experimental implications. As the recent QENS measurements on NaCrO$_2$ was performed on the powder sample, the measured intensity is not the contribution from a single particular $\bm{q}$, but rather the average over the contributions from various $\bm{q}$'s. Our result suggests that, through the powder averaging over various $\bm{q}$'s, the resulting QE scattering intensity would become a superposition of many Lorentzians with continuously distributed widths. 
Since the free $Z_2$ vortex near $T_V$ yields the central peak of the narrowest width of $\sim 0.001J$ with a significant intensity among nearby $\bm{q}$'s, it would be visible even after the powder average. Indeed, the recent QENS experiment succeeded in clearly catching the corresponding sharp QE scattering of its width 0.01meV $\sim$ 0.001$J$ \cite{Tomiyasu}.

 We note that, in the analysis of Ref.[\citen{Tomiyasu}], the superpositions of these continuously distributed Lorentzians are approximated by the three Lorentzians, i.e., very sharp $L_1$-component of $\sim$0.01meV width describing the free $Z_2$ vortex, $L_2$-component of $\sim$0.1meV width describing the typical $Z_2$-vortex pairs, and $L_3$-component of $\sim$1meV width describing the damped spin waves. The analysis, though an approximation, is expected to capture the essential part of relevant fluctuations. Our present analysis then suggests that the $L_1$-component is borne by the contributions close to the $K$-point, while the $L_2$-component borne by those away from the $K$-point.
Meanwhile, if the QENS experiment could be performed on the single crystal, more detailed information including the $\bm{q}$ dependence might well be obtained, which would make further detailed comparison with the theory possible. Thus, single-crystal QENS measurements on the $Z_2$-vortex-bearing magnet remain to be an interesting future task.

We are thankful to ISSP, the University of Tokyo, for providing us with CPU time. This study was financially supported by JSPS KAKENHI (JP17H06137, JP18K03503). 


\end{document}